\documentclass[12pt]{article}
\usepackage{aat}

\def\msun{M_\odot}

\def\etal{{\it et.~al.~}}
\def\eg{{\it e.g.~}}
\def\ltsima{$\; \buildrel < \over \sim \;$}
\def\simlt{\lower.5ex\hbox{\ltsima}}
\def\gtsima{$\; \buildrel > \over \sim \;$}
\def\simgt{\lower.5ex\hbox{\gtsima}}
\begin{document}

\mylabel{}

\mytitle{Metal enrichment of the high-z intergalactic medium}

\myauthor{Yu. A. Shchekinov}

\myadress{Department of Physics, Rostov State University, \\ 
344090 Rostov on Don, Russia}
\mydate{}

\myabstract{Recent observations show the presence of metals in low-density 
Ly$\alpha$ forest absorbers at high redshift ($z\sim 3$). It remains still 
far from being clearly understood what mechanisms spread metals over Mpc scales 
from the parent galaxies, whether metals are homogeneously distributed in 
the intergalactic medium (IGM), how metallicity of the IGM does evolve. 
These questions are brifely addressed in this review.  
}

\section{Introduction}

During the last decade it became clear owing to cosmological hydrodynamic
simulations that the Ly$\alpha$ forest is a result of the growth of 
large scale structure in the universe in the presence of UV field [1--4].  
In this scenario the baryons in the universe form a network structure 
where most of gas and galaxies are accumulated in narrow and dense walls 
of the web, while the most of volume is filled with gas of lower density 
(close or below the mean cosmic baryon density). These underdense regions 
are known as voids and represent the truly intergalactic medium, which is 
seen in the Ly$\alpha$ forest lines of the low column density end, 
$\log N(HI)<14.5$ cm$^{-2}$.  

First detections of metals in high-redshift ($z\sim 3$) Ly$\alpha$ forest 
systems with metallicity $\rm [C/H]\simeq -2.5\pm 1$ were related 
basically to the intermediate range of column densities, $\log N(HI)\geq 
14.5$ (cm$^{-2}$) [5, 6]. In 1998 two independent groups observed Ly$\alpha$ 
forest with lower column densities [$\log N(HI)<14.5$], and using different 
observational techniques have reached controversary conclusions: Cowie \& 
Songaila [7] with pixel-to-pixel optical depth observations reported 
on metal detection with $\rm [C/H]=-2.5$ at $z\simeq 3$ for $\log N(HI)=13.5$; 
Lu \etal [8] with a stacking observations were only able to reach an 
upper limit $\rm [C/H]<-3.5$ for $\log N(HI)<14$. In [9] the two techniques 
were critically analyzed and both found to suffer limitations. It was 
concluded, however, that the stacking is stronger subjected to smearing out 
due to a random redshift offset between the Ly$\alpha$ and CIV lines, and 
thus the amount of metals is underestimated. Instead, they found the 
pixel-to-pixel optical depth procedure to be more robust against the redshift 
offset, and the abundances obtained with using this technique more 
confident. Applying pixel-to-pixel measurements for the currently most 
sensitive Keck/HIRES spectra of QSO APM 08279+5255 ($z=3.87$) [9] and 
Q1422+231 ($z=3.625$) [10] the abundance of carbon in low-density 
($\log N(HI)< 14.5$) Ly$\alpha$ forest was found as $\rm [C/H]=-2.6$. 
Observations in OVI lines allow detection of absorptions from HI optical 
depths of one order of magnitude lower than CIV absorptions do, and thus 
can probe less dense Ly$\alpha$ gas. The first detection of OVI absorptions 
at $z=2-3$ from underdense regions ($\rho/\langle \rho\rangle \leq 1$, which 
represent thus the true IGM) was reported in [11]. Very 
recently Songaila [12] has examined the presence of metals at highest 
redshift, and concluded that the column density distribution of CIV remains 
invariant up to redshift $z=5.5$ where Ly$\alpha$ forest becomes thick, 
and the metallicity at this redshift exceeds $[Z]\simeq -3.5$. The invariancy 
of the CIV density distribution may indicate that analogously to lower $z$ 
the metallicity of low density regions lies in same range. The presence of 
metals at high redshift gives direct evidences of stellar activity at early 
epochs in the universe corresponding to redshifts as high as at least $z>5.5$. 
The question of a primary importance here is, however, how the metals were  
spread over Mpc scales in the IGM very far from the galaxies where they 
have been synthesized. The seriousness of this question becomes obvious if 
we formally devide the characteristic size of typical void of several Mpc over 
the Hubble time $t_H(z)$: at $z=3$ this is $\sim 500$ km s$^{-1}$.

\section{Pop III objects and supernovae driven galactic outflows}

Two episodes of metal enrichment of the Ly$\alpha$ forest absorbers 
are being discussed: early pre-enrichment by a widespread initial star 
formation, normally attributed to Pop III objects [13--16], and local 
enrichment at later stages either by star formation within the clouds 
themselves or by contamination from nearby galaxies [17]. The estimated 
amount of metals produced by the Pop III objects (presumably the first 
galaxies of small masses) is sufficient to pollute the IGM to an average 
metallicity from  $\sim 10^{-4}Z_\odot$ [13--15] to $\sim 0.003~Z_\odot$ 
[16]. The expected spatial distribution of metals is apparently very 
putchy, and strongly depends on the mixing processes. The amount of metals 
in a hot cavity produced by the exploded SNe in Pop III objects can reach 
$0.2-0.4~Z_\odot$ [16] or even $\sim Z_\odot$ [18]. If however, the metals 
are mixed with the swept up shell, the resultant metallicity decreases 
by two orders of magnitudes to the level shown above. In general, the gas 
metallicity is $Z=M_Z/V_Z$, where $M_Z$ is the total mass of metals ejected 
in the volume $V_Z=4\pi R_e^3/3$, $R_e$ is determined from the condition that 
the blowing out flow is confined by the IGM pressure 

\be
R_e=\sqrt{{\dot M_e v_e\over 4\pi P_{\rm igm}}},
\ee
$\dot M_e$ is the mass ejection rate, $v_e$, 
the velocity of the outflow. 
Assuming $\dot M_e$ to be a fraction $\xi\sim 0.1$ of the star formation rate 
$\dot M_\ast\propto M_h$ [19], and $M_Z=\dot M_e\tau$ with $\tau$ being the 
period of star formation activity, and substituting for $v_e$ the escape 
velocity for a dark matter halo of mass $M_h$ and radius $r_h$: 
$v_e=\sqrt{G M_h/r_h}$, one arrives at the metallicity inversely proportional 
to galactic mass $Z\propto \tau (fM_h)^{-1}$, where $f=f(M_h)$ is the fraction 
of mechanical energy of the galaxy which can blow out, $f\simeq 1$ for 
galaxies of small masses $\Omega_bM_h<10^9(1+z)^{-3/2}\msun$, and 
decreases for larger masses [18]. It is readily seen that even if the metals 
are mixed inside the blown out spheres, their spatial distribution in the 
IGM is highly non-homogeneous varying from one parent galaxy to the other. 
Thus, even in those cases when the volume filling factor of the blown out 
spheres is $q\sim 1$ and they are partilly overlapped (see, \eg [16]), 
an efficient mixing mechanism must be present in the early univers in order 
that the metals were distributed homogeneously. Note, however, that metal 
enriched gas loses its energy radiatively very easily, forming due to thermal 
instability dense condensations, and thus preventing efficient mixing on 
smaller scales [20].  

For low mass galaxies a few SNe explosions taking place in an OB-association 
of a moderate mass can be enough to drive a blowout. The situation changes 
in large galaxies, where due to increasing sizes and potential well ejection 
of shock processed gas becomes more difficult, so that most of metals remains 
confined to the galaxy. The fraction of mechanical energy $f$ of the galaxy 
which can drive a blowout is [18] 

\be
f={\ln(N_M/N_c)\over \ln(N_M/N_m)},
\ee
where $N_M$ ($N_m$) is the maximum (minimum) possible number of SNe in  
OB-associations, $N_c=t_{ob}L_c/\epsilon_0$ is the critical number of SNe 
corresponding to the 
critical mechanical luminosity required for a blowout to occur (see [18])

\be
L_c\sim 10^{14}c_h^2\Omega_b\left({M_h\over \msun}\right)(1+z)^{3/2}
~{\rm erg~s^{-1}},
\ee 
$c_h$ is the sound speed in gaseous galactic halo, 
$t_{ob}\sim 30$ Myr, the life time for the 
lowest SNe progenitor mass, $\epsilon_0=10^{51}$ erg. Thus, 
$f$ vanishes logarithmically when $N_c$ is approaching $N_M$. Assuming 
that $N_M$ does not depend on galactic mass, one can obtain the critical 
galactic mass above which blowout events are inhibited. For $N_M=10^3$ and 
$c_h\sim 100$ km s$^{-1}$ the critical mass is $\Omega_bM_h\sim 
10^{11}(1+z)^{-3/2}\msun$, meaning that more massive galaxies do not 
contribute to the enrichment of the IGM and the main pollutors are the 
low mass galaxies.   

The volume filling factor of the regions containing the blown out material 
remains always much less than one, $q\sim 10^{-4}$ [18], only if most of the 
galaxies experienced a starburst regime the corresponding filling factor 
may reach $q\simgt 0.2$ [16]. Thus, if metals in the IGM are distributed 
relatively homogeneously, one has to assume that efficient mixing mechanisms 
were acting in the early universe. In this connection, it is interesting to 
note that observations of the local Ly$\alpha$ forest place limits on  
metal spread within 150-200 kpc from the galaxy [21]. 

\section{Turbulent mixing}

It was suggested in [17] that metals are transported predominantly by 
violent gas flows following galactic mergings. This idea was further developed 
in [18] where peculiar motions of galaxies and subgalactic halos were shown 
to power efficiently turbulent flows of the ejected gas. The diffusion 
coefficient needed for an efficient mixing can be estimated by requiring that 
in proper Hubble time $t_H(z)$ turbulence spreads metals over the 
characteristic distance between galaxies
$\sim 150h^{-1}(1+z)^{-1}$ kpc [16], where $h=H_0/100$ km s$^{-1}$ 
Mpc$^{-1}$: $\kappa\sim 6\times 10^{29}h^{-1}(1+z)^{-1/2}$ cm$^2$ s$^{-1}$. 
In [18] the turbulent diffusion coefficient was estimated approximately as 
$\kappa_t\sim \Omega_b^{-1}c_s^2t_J$, where $c_s$ is the sound speed in the 
IGM, $t_J=(N_h\lambda_J^2\sigma)^{-1}$, the characteristic time interval 
between subsequent encounters of the halos onto a Jeans sphere in 
the IGM, $\lambda_J$ is the corresponding Jeans length, $N_h$, the 
number of dark halos in unit volume, $\sigma$ is their velocity dispersion. 
For the IGM temperature $T\simeq 2\times 10^4$ K, and a comoving number 
density of dark halos of $70h^3$ Mpc$^{-3}$ (corresponding approximately 
to $M_h\sim 10^8h^{-1}\msun$) [16], 
one can estimate $\kappa_t\sim 4\times 10^{29}h^{-1}m^{-1}$ cm$^2$ s$^{-1}$, 
$m$ is Mach number of halos in the IGM. As a consequence, the turbulent 
zones are shown to get partially overlaped at $z\sim 1$, and reach the 
porosity $q\simeq 1$ at $z\sim 0$ [18]. However, metals still remain 
distributed rather inhomogeneously, only at $z\sim 0$ the scatter is 
0.03 around the mean value $0.1~Z_\odot$, while at $z\sim 3$ it is more 
than two orders of magnitude. Thus, even under such a strong diffusivity 
the enriched regions keep the identity of the parent galaxy produced pollution.   

\section{Radiation pressure}

Dust particles ejected from galaxies by radiation pressure can be one of 
the possible sources of metals in the IGM. This possibility was first 
discussed in [22] with a negative conclusion. Detailed study initiated in 
[23] showed instead that the galactic radiation field is sufficient enough 
to expell dust particles in a relatively short time $\sim 10^8$ yr at high 
distances from the plane, practically into the intergalactic space. In [24] 
dynamics of radiatively ejected dust were considered with incorporation 
the effects of grain destruction. The characteristic column density $N(H)$ 
needed for destruction of a grain with radius $a$ is $N_d(H)\simeq 
4\rho a/Y_sm_T$, where $\rho=3$ g cm$^{-3}$ is the grain density, $Y_s$ is 
the angle-averaged sputtering yield, $m_T$, the target mass. For a fast 
moving grain, $v_g>100$ km s$^{-1}$, $Y_s\sim 0.1$ in collisions with helium, 
and $Y_s\sim 0.01$ in collisions with hydrogen [25, 26]. Substituting here 
$m_T\sim 20m_H$ one obtains $N_d(H)\sim 3\times 
10^{20}(a/0.1~\mu{\rm m})$ cm$^{-2}$, which determines a lower limit for 
the grain sizes to be survived: the vertical galactic column density should 
be smaller than the critical value, $N(H)<N_d(H)$. The critical column 
density $N_d(H)$ for $a=0.1~\mu$m 
is practically coincident with the maximal vertical column densites of hydrogen
and electrons in the Milky Way. Fast particles of smaller sizes can survive 
only if they start at $z\geq 500$ pc above the plane. One should stress, 
however, that at lower distances from the plane, $z<200$, dust particles 
move  much slower, $v_g\ll 100$ km s$^{-1}$ [27], so that $Y_s$ drops sharply 
by several orders of magnitude. 

For a destroying grain the extinction and radiation pressure decrease as $a$, 
resulting in decrease of the velocity of the ejected particle: for graphite 
and silicate particles of the initial radius $a=0.1~\mu$m, starting at $z=1$ 
kpc above the galactic plane, their asymptotic velocities 
at time $\sim 100$ Myr drop from $\sim 1600$ to $800$ km s$^{-1}$, and 
from $\sim 700$ to $\sim 350$ km s$^{-1}$, respectively, when dust 
destruction is taken into consideration [24]. However, after $t\sim 1$ Gyr 
the corresponding distances can reach $\sim 1$ Mpc for graphites, and 
$\sim 0.7$ Mpc for silicates. Thus, at redshift $z\sim 3$ radiatively driven 
$0.1~\mu$m dust particles can, in principle, transport metals from galaxies 
into the IGM over Mpc scales. Smaller particles, though, are strongly 
confined to galactic disks mainly by collisional drag, so that both graphite 
and silicate grains with the initial radius $a=0.01~\mu$m can reach in 1 Gyr 
only $\sim 20-30$ kpc from the disks [24]. The resulting steady distribution 
of the density of dust particles is highly nonuniform, with $\rho_d\sim 
10^{-30}$ g cm$^{-3}$ at $r \sim 30-60$ kpc, and $\rho_d\sim 10^{-32}$ 
at $r\sim 120$ kpc, which is sharper than $r^{-2}$. 

Therefore, only large particles can be efficient in spread of metals in the 
IGM. For a galaxy with radiation flux $G$ in Habing units ($1.6\times 10^{-3}$ 
erg cm$^{-2}$ s$^{-1}$), and gravitational acceleration $g_9$ (in $10^{-9}$ 
cm s$^{-2}$), only grains with $a<0.2~Gg_9^{-1}~\mu$m can be radiatively 
ejected. Assuming $g\propto M/R^2$ and $G\propto M$, one can find for 
galaxies satisfying a Tully-Fisher relation with $R\propto M^{1/3}$  
that only grains of radius $a<0.1 (M/M_{\rm MW})^{2/3}\mu$m can be ejected, 
$M_{\rm MW}$ is the mass of the Milky Way galaxy. As soon as the sizes of 
dust particles are restricted from below in order that they were survived 
when being ejected, one can conclude that only massive galaxies are 
efficient in pollution of the IGM with metals. Note however, that if at 
smaller distances from the plane dust particles are transported by slow 
hydrodynamic motions, such that the sputtering yield remains $Y_s\ll 0.01$, 
and only at higher distances, $z\simgt 500$ pc, they are accelerated by 
radiation pressure to high velocities, $v_g>100$ km s$^{-1}$, the lower 
limit can be as small as $a<10$\AA. At such conditions a still considerable 
mass fraction of dust (for a standard size distribution) can be expelled into 
the IGM. Quite recently this mechanism was explored in framework of large 
scale cosmological simulations [28]. It is worth noting, that spatial 
distribution of dust (and as a consequence, metals in gas phase) is 
demonstrated in these simulations to be highly 
inhomogeneous: at $z=3$ low density regions [with $N(HI)<10^{15}$ cm$^{-2}$] 
show metallicities in the range $Z=10^{-5}-0.1$ (absolute values). It is 
quite similar to the predictions of [18] based on a model of turbulent 
mixing. 
 
Regular magnetic field in the insterstellar medium can change the picture 
crucially. Dust grains in diffuse intercloud gas exposed to the ionizing 
radiation field are positively charged up to $\sim 200~e$ [29]. The 
corresponding gyro-radius of a grain is $r_g=m_gcv_g/Z_geB\sim 3\times 
10^{15}$ cm for $a=0.1~\mu$m, $v_g=10^6$ cm s$^{-1}$ (typical velocity 
dispersion in the intercloud gas), and $B=1~\mu$G. It was suggested in 
[30] that due to fluctuations of charging processes dust grains spend 
considerable time intervals in a neutral state. However, characteristic 
charging times for a grain with $a\sim 0.1~\mu$m are very short: 
$\sim 10^2-10^3$ s, for charging due to photoemission of electrons and 
recombination of electrons on the grain, so that the equilibrium between the 
two processes provides the grain charge of the order of 200 $e$. At these 
conditions, the probability of a fluctuation which can neutralize such a high 
grain charge is negligible. It is obvious that even in dwarf galaxies where 
magnetic field seems to be weaker, the gyro-radius remains much smaller 
than characteristic galactic scales. Note however, that recent polarimetric 
observations show that dwarf galaxies can posses a sufficiently strong 
magnetic field: $\sim 8\mu$G in NGC 4449 [31]. This means that dust particles 
are strongly confined to magnetic field, and thus can be ejected out of 
galactic disks only in those regions where magnetic field has a predominantly 
vertical component. In spiral galaxies (and in those dwarfs with sufficiently 
strong magnetic field) such a vertical magnetic field can 
form due to the Parker instability, which in turn can be initiated easily by 
SNe explosions with the total explosion energy much less than required for 
a blowout [32, 33]. At latest stages (typically 100 Myr) growing magnetic 
loops seem to connect to the infinity, and thus provide corridors for 
photoejection of dust. Under certain conditions the Parker instability can be 
initiated by radiation pressure acting on dust particles, as mentioned 
in [30] and recently investigated numerically in [34]. 
 
The effects of magnetic field are also important in those cases when dust 
particles are confined in clouds or clumps, for which the charge to mass ratio 
is zero, and correspondingly $r_g=\infty$. In this case the proper magnetic 
field of clouds is connected to  the galactic magnetic field, so that magnetic 
tension $({\bf B\cdot\nabla}){\bf B}/4\pi\simgt 3\times 
10^{-33}B^2_{\mu{\rm G}}$ [for $|\nabla|\sim (100~{\rm pc})^{-1}$] is 
always larger than momentum transfered from radiation field 
$\sim 3\times 10^{-36}Gn$, where $G$ is in Habing units, and the abundance of 
dust is taken $10^{-12}$.  

\section{Conclusions}

Detection of metals in low column density Ly$\alpha$ forest absorbers seem 
to imply that efficient ejection and transport mechanisms are at work at 
high redshift. The nature of these mechanisms is not yet clearly understood. 
Nevertheless, recent studies have largely enlightened the problem, and 
can be summarized as:

1. At early stages metals ejected into the IGM are confined relatively close 
to the pollutant galaxies by the IGM pressure. The volume filling factor can 
be large only when the star formation efficiency is high, corresponding to a 
starburst mode. However, even in this case spatial distribution of metals is 
highly inhomogeneous -- inversly proportional to the mass of a pollutant 
galaxy. 

2. In order to make the distribution more smooth and widespread, efficient 
mixing mechanisms must have been acting. Of these mechanisms turbulent mixing 
by tidal interactions of galaxies may be quite a powerful, resulting in a 
large volume filling factor of metal enriched IGM. In this case the spread 
in metallicities is smaller, but still can be as large as two orders of 
magnitude at $z\sim 3$, and about $\pm 0.03~Z_\odot$ at $z=0$. In both these 
cases though, an enhanced radiative cooling of the metal enriched gas 
initiates growth of dense condensations through thermal instability, which 
prevents smooth redistribution of metals, thus exacerbating the mixing problem. 

3. Similar distribution is expected also if the IGM is enriched through 
radiatively driven ejection of dust from galaxies. However, not all is clear 
here. In particular, such questions as: what fraction of dust particles can be 
expelled into the IGM from deep regions of the galactic disks, with 
$N(H)>N_d(H)$ where they are mainly produced; how galactic magnetic field 
affects this process; how the ejection efficiency varies with redshift -- are 
still to be answered.  

\vskip 0.5cm
\noindent
{\it Acknowledgements} \\
\vskip 0.1 cm
\noindent
This work was supported by RFBR (grant N 00-02-17689) and by Federal Programme 
``Astronomy'' (grant N 1.3.1.1). 
\vskip 0.8 cm
\noindent
{\it References}
\vskip 0.1 cm
\noindent

\rf [1] Petitjean, P., M\"ucket, J. P., Kates, R. E. (1995) {\it 
Astron. \& Astrophys.}, {\bf 295}, L9

\rf [2] Hernquist, L., Katz, N., Weinberg, D. H., Miralda-Escud\'e, J. 
(1996) {\it Astrophys. J.}, {\bf 457}, L51

\rf [3] Zhang, Y., Meiksin, A., Anninos, P., Norman, M. L. (1998) 
{\it Astrophys. J.}, {\bf 495}, 63

\rf [4] Rauch, M. (1998) {\it Ann. Rev. Astron. \& Astrophys.}, {\bf 
36}, 267

\rf [5] Cowie, L. L., Songaila, A., Kim, T., Hu, E. M. (1995) 
{\it Astron. J.}, {\bf 109}, 1522

\rf [6] Tytler, D., Fan, X. M., Burles, S., Cottrell, L., Davis, C., 
Kirkman, D., Zuo, L. (1995) in: {\it QSOs Absorption Lines}, Proc. 
ESO Workshop, ed. Meylan, G. (Heidelberg: Springer) 

\rf [7] Cowie, L. L., Songaila, A. (1998) {\it Nature}, {\bf 394}, 44

\rf [8] Lu, L., Sargent, W. L. W., Barlow, T. A., Rauch, M. (1998) 
preprint, astro-ph/9802189

\rf [9] Ellison, S. L., Lewis, G. F., Pettini, M., Sargent, L. W. L., 
Chaffee, F. M., Irwin, M. J. (1999) {\it Astrophys. J.}, {\bf 520}, 456

\rf [10] Ellison, S. L., Songaila, A., Schaye, J., Pettini, M. (2000) in:
{\it Cosmic Evolution and Galaxy Formation: Structure, Interactions and 
Feedback}, ASP Conf. Series, J. Franco, E. Terlevich, O. L\'opez-Cruz, 
I. Aretxaga, eds., V. , p. 

\rf [11] Schaye, J., Rauch, M., Sargent, W. L. W., Kim, T.-S. (2000) 
{\it Astrophys. J.}, {\bf 541}, L1

\rf [12] Songaila, A. (2001) {\it Astrophys. J.}, {\bf 561}, L153

\rf [13] Gnedin, N. Y., Ostriker, J. P. (1997) {\it Astrophys. J.}, 
{\bf 486}, 581

\rf [14] Miralda-Escud\`e, J., Rees, M. J. (1997) {\it Astrophys. J.}, {\bf 
478}, L57

\rf [15] Nath, B., Trentham, N. (1997) {\it Month. Not. Roy. Astron. Soc.}, 
{\bf 291}, 505  

\rf [16] Madau, P., Ferrara, A., Rees, M. J.  (2001) {\it Astrophys. J.}
{\bf 555}, 92

\rf [17] Gnedin, N. Y. (1998) {\it Month. Not. Roy. Astron. Soc.}, {\bf 294}, 
407

\rf [18] Ferrara, A., Pettini, M., Shchekinov, Yu. A. (2000) {\it Month. 
Not. Roy. Astron. Soc.}, {\bf 319}, 539

\rf [19] Ferrara, A., Tolstoy, E. (2000) {\it Month. Not. Roy. Astron. 
Soc.}, {\bf 313}, 291

\rf [20] Dedikov, S., Shchekinov, Yu. A. (2002) {\it Astr. Rept.}, submitted

\rf [21] Penton, S. V., Stocke, J. T., Shull, J. M. (2002) {\it 
Astrophys. J.}, {\bf 565}, 720

\rf [22] Pecker, J. C. (1972) {\it Astron. and Astrophys.}, {\bf 18}, 253

\rf [23] Ferrara, A., Ferrini, F., Franco, J., Barsella, B. (1991) 
{\it Astrophys. J.}, {\bf 381}, 137

\rf [24] Shustov, B. M., Vibe, D. Z. (1995) {\it Astron. Rept.}, {\bf 39}, 
578

\rf [25] Draine, B. T., Salpeter, E. E., (1979) {\it Astrophys. J.}, 
{\bf 231}, 77

\rf [26] Draine, B. T., (1995) {\it Astrophys. Space Sci.}, {\bf 233}, 111 

\rf [27] Dettmar, R.-J., Schr\"oer, A., Shchekinov, Yu. A. (1999) in: {\it 
Proc. 26th ICRC}, p. 298

\rf [28] Aguirre, A., Hernquist, L., Katz, N., Gardner, J., Weinberg, D. 
(2001) {\it Astrophys. J.}, {\bf 556}, L11

\rf [29] Ferrara, A., Dettmar, R.-J. (1994) {\it Astrophys. J.}, {\bf 427}, 
155

\rf [30] Ferrara, A. (1998) in: {\it The Local Bubble and Beyond}, 
D. Breitschwerdt, M. J. Freyberg, J. Tr\"umper, eds., Lecture Notes 
in Physics, {\bf 506}, p. 371

\rf [31] Chyzy, K. T., Beck, R., Kohle, S., Klein, U., Urbanik, M. 
(2000) {\it Astron. Astrophys.}, {\bf 355}, 128; {\bf 356}, 757

\rf [32] Kamaya, H., Mineshige, S., Shibata, K., Matsumoto, R. (1996) 
{\it Astrophys. J.}, {\bf 458}, L25

\rf [33] Steinecker, A., Shchekinov, Yu. A. (2001) {\it Month. Not. 
Roy. Astron. Soc.}, {\bf 325}, 208

\rf [34] Kopp, A., Shchekinov, Yu. A. (2002) in preparation

\end{document}